\documentclass{article}
\usepackage[utf8]{inputenc}
\usepackage{spconf,amsmath,graphicx}
\usepackage{cite}
\usepackage{times}
\usepackage{latexsym}
\usepackage{color}
\usepackage{listings}
\usepackage{textcomp}
\usepackage{caption}
\usepackage{mathtools}
\usepackage{wrapfig}
\usepackage{multirow}
\usepackage[hyphens]{url}
\usepackage[hidelinks]{hyperref}
\usepackage{cleveref}
\usepackage{amsmath,graphicx}
\usepackage{adjustbox}
\usepackage{siunitx}
\usepackage{setspace}
\usepackage{caption}
\usepackage{enumitem}
\usepackage{etoolbox}

\renewrobustcmd{\bfseries}{\fontseries{b}\selectfont}
\renewrobustcmd{\boldmath}{}
\newrobustcmd{\B}{\bfseries}

\newsavebox\CBox

\makeatletter

\renewcommand{\section}{\@startsection
   {section}%
   {1}%
   {}%
   {-0.4\baselineskip}%
   {0.2\baselineskip}%
   {}}%

\renewcommand{\subsection}{\@startsection
  {subsection}%
  {2}%
  {}%
  {-0.1\baselineskip}%
  {0.1\baselineskip}%
  {}}%

\renewcommand{\subsubsection}{\@startsection
  {subsubsection}%
  {3}%
  {}%
  {-0.2\baselineskip}%
  {0.2\baselineskip}%
  {}}%

\g@addto@macro\normalsize{%
  \setlength\abovedisplayskip{3pt plus 2pt minus 1pt}
  \setlength\belowdisplayskip{3pt plus 2pt minus 1pt}
  \setlength\abovedisplayshortskip{2pt plus 2pt minus 1pt}
  \setlength\belowdisplayshortskip{2pt plus 2pt minus 1pt}
}

\setlist{
        itemsep=0pt,
        parsep=1pt plus 1pt minus 1pt,
        topsep=1pt plus 1pt minus 1pt,
        partopsep=0pt}

\makeatother

\DeclareUnicodeCharacter{00A0}{}

\title{A systematic comparison of grapheme-based vs.~phoneme-based label units for encoder-decoder-attention models}
\name{Mohammad Zeineldeen, Albert Zeyer, Wei Zhou, Thomas Ng, Ralf Schlüter, Hermann Ney}
\address{
  Human Language Technology and Pattern Recognition,
  Computer Science Department, \\
  RWTH Aachen University, 52062 Aachen, Germany, \\
  AppTek GmbH, 52062 Aachen, Germany \\
  {\strut \small \tt
    \{zeineldeen, zeyer, zhou, schlueter, ney\}@cs.rwth-aachen.de, thomas.ng@rwth-aachen.de
  }
}

\begin{document}

\maketitle
%
% The total length of the abstract is limited to 200 words.
\begin{abstract}
% Common end-to-end models like CTC
% or encoder-decoder-attention models
% use characters or subword units like BPE as the output labels.
% %While characters or subwords are simple,
% %the model cannot easily recognize words with uncommon pronunciations
% %not seen during training,
% %and there is no easy way to adapt the model without retraining it.
% %Phonemes on the output label would allow for
% %easy adaptation and extension of the lexicon.
% We do systematic comparisons
% between variants of grapheme-based and phoneme-based output labels
% with an encoder-decoder-attention model.
% These can be single phonemes ($\approx$ 47 labels),
% or multiple phonemes together in one output label,
% such that we get phoneme-based subwords.
% For this purpose, we introduce phoneme-based BPE labels.
% In further experiments,
% we extend the phoneme set
% by auxiliary units
% to be able to discriminate homophones. %(different words with same pronunciation).
% This enables a simplified and efficient decoding algorithm.
% We perform experiments on Switchboard 300h and LibriSpeech 960h
% and we can show that our phoneme-based models can be competitive
% to the grapheme-based models.
%
Following the rationale of end-to-end modeling,
CTC, \mbox{RNN-T} or encoder-decoder-attention models
for automatic speech recognition (ASR)
use graphemes or grapheme-based subword units based on e.g.~byte-pair encoding (BPE).
The mapping from pronunciation to spelling % in these cases
is learned completely from data.
%which might suffer when only limited training resources are available.
In contrast to this,
classical approaches to ASR employ secondary knowledge sources
in the form of phoneme lists to define phonetic output labels
%for acoustic modeling
and pronunciation lexica. %to perform the mapping from pronunciations to words.
In this work, we do a systematic comparison
between grapheme- and phoneme-based output labels
for an encoder-decoder-attention ASR model.
We investigate the use of single phonemes as well as BPE-based
phoneme groups as output labels of our model.
To preserve a simplified and efficient decoder design,
we also extend the phoneme set by auxiliary units
to be able to distinguish homophones.
Experiments performed on the Switchboard 300h and
LibriSpeech benchmarks show that phoneme-based modeling is competitive to
grapheme-based encoder-decoder-attention modeling.
\end{abstract}

\section{Introduction \& Related Work}

End-to-end models such as encoder-decoder-attention models
\cite{%sutskever2014seq2seq,
bahdanau2015align,luong2015attentionmt,
%chorowski2015attention,chan2016las,%
zeyer2018:asr-attention,park2019specaugment,%
zeyer2019:trafo-vs-lstm-asr,tuske2020swbatt},
connectionist temporal classification (CTC) models
\cite{graves2006connectionist,miao2015eesen,%
amodei2015deepspeech2,zweig2017allneural},
or the recurrent neural network transducer (RNN-T)
\cite{graves2012seqtransduction,
%prabhavalkar2017comparison,rao2017rnnt,%
%jain2019rnnt,li2019rnnt,han2020contextnet
zeyer2020:transducer}
have shown competitive performance for speech recognition.
%while being relatively simple
End-to-end models usually operate on
\begin{itemize}
\item \emph{graphemes} (characters),
\item \emph{graph.~based subword units}
(\emph{byte-pair encoding (BPE)} \cite{gage1994bpe,sennrich2015neuralbpe,zeyer2018:asr-attention},
%WordPieces /
word piece model (WPM) \cite{schuster2012wordpieces,wu2016googlenmt}, %,park2019specaugment},
unigram language model (LM) based segmentation
\cite{kudo2018subword,bostrom2020unigramlm},
latent sequence decompositions \cite{chan2016lsd,drexler2020subword},
or pronunciation-assisted char.~subwords \cite{xu2019pronsubword})
\item or \emph{whole words}
\cite{soltau2017wordctc,audhkhasi2017wordctc,%
palaskar2018word,chen2018modular,audhkhasi2018building,das2019wordctc}.
\end{itemize}

% Graphemes and subwords have the advantage of allowing out-of-vocabulary words
% and simplicity because no pronunciation lexicon is needed.
% %End-to-end models usually do not operate on phonemes.
% %In this work, we study the effect of different label units on encoder-decoder-attention models.
% Having an explicit pronunciation lexicon and phoneme labels
% allow to easily adapt some pronunciation,
% or extend by new words with new uncommon pronunciations
% \cite{bruguier2019phoebe}.
% %This is a common problem for grapheme or subword based models
% %in case of uncommon pronunciations \cite{bruguier2019phoebe}.
% This is \emph{tradeoff} between simplicity
% towards greater flexibility and controllability
% of the pronunciations.

Grapheme-based modeling bears the advantages of easily allowing to recognize
out-of-vocabulary words, enabling modeling, training and decoding simplicity
and avoiding potential
errors introduced by secondary knowledge sources like pronunciation lexica.
The disadvantage of grapheme-based modeling might be limited scalability to
low-resource situations.
Phoneme-based modeling %using explicit pronunciation lexica
on the other hand easily enables adaptation to new pronunciations or
inclusion of words with uncommon pronunciations \cite{bruguier2019phoebe}.
%while potentially suffering from pronunciation errors.
These output-label modeling alternatives
therefore provide a tradeoff between simplicity, flexibility,
and controllability.

Hybrid neural network (NN) - hidden Markov models (HMMs)
\cite{bourlard1994hybrid,robinson1994rnnhmm}
usually operate on context-dependent phonemes
\cite{zeyer17:lstm,zeineldeen:icassp20},
although it has been shown that they can also work
on context-dependent graphemes
\cite{kanthak:icassp2002,killer2003grapheme,sung2009graphemes,le2019senones}.
%As these are too much labels,
Due to the resulting combinatorial complexity, phonemes (graphemes)
in context usually are clustered
via classification and regression trees (CART)
\cite{breiman1984cart,young1992cart}.

All end-to-end models % like attention models
easily allow for phoneme-based labels as well,
although in most cases that requires a more complex decoder
like a weighted finite-state transducer (WFST) decoder \cite{irie19:phoneme},
and a pronunciation lexicon.
Phoneme-based CTC models
(context-independent or clustered context-dependent)
\cite{sak2015fast,miao2016ctc,chen2018modular}
usually perform better than grapheme-based CTC models,
just as hybrid NN-HMMs.
Phoneme-based encoder-decoder-attention models
\cite{sainath2018nolex,zhou2018comparison,%
irie19:phoneme,wang2020phonebpe,kubo2020joint}
have shown mixed results so far
-- in most cases the performance
was slightly worse than grapheme-based attention models,
or only the combination of both helped.
More recently also phoneme-based \mbox{RNN-T}-like models
\cite{hu2019phoneme,variani2020hat}
were studied,
although it is unclear
whether pure phoneme-based \mbox{RNN-T} models
perform better than pure grapheme-based models.
%\TODO Kazuki phone \cite{irie19:phoneme}:
%\begin{itemize}
%	\item Encoder-decoder-attention models on 960hr Librispeech dataset
%	\item Comparison between single grapheme and single phoneme
%	\item Unknown token (UNK) as part of phoneme vocab.
%	\item End-of-word (EOW) as part of phoneme vocab.~to model word boundary
%	\item Decoding with WFST decoder \& word-based n-gram LM
%	\item 70 phone units
%	\item Performance on phonemes is worse than graphemes (\TODO quantify)
%	\item No phoneme-BPE but only single phonemes.
%\end{itemize}

% \cite{wang2020phonebpe} so far is the only paper studied phon-bpe
Subwords based on phonemes, like phoneme-BPE,
was only studied so far by \cite{wang2020phonebpe},
which is a hybrid CTC / attention model \cite{watanabe2017ctcatt}.
They use multiple LMs in decoding: a phoneme-BPE LM and a word LM.

%\TODO Single grapheme and grapheme-based subwords (char-BPE or word pieces)
%has been compared
%before for \mbox{RNN-T} \cite{rao2017rnnt}.

In this work, we perform a systematic comparison of grapheme-based and
phoneme-based output labels with an encoder-decoder-attention model.

\section{Variation of label units}

We want to perform a systematic study of the difference
between phonemes, graphemes,
subwords based on phonemes or graphemes,
and whole words.
In this work, we focus on attention-based encoder-decoder models.
We note that the results of such label unit study
will likely look different depending on the type of model.
E.g.~subwords and words can possibly have variations
in their observed audio lengths,
which might be less of a problem for label-synchronous models
but more a problem for time-synchronous models.
%E.g.~maybe the optimal label unit for hybrid NN-HMMs or CTC models are phonemes,
%while the optimal label unit for label-synchronous models are subwords.
For hybrid NN-HMMs or CTC models,
we also know that clustered context-dependent labels help a lot
\cite{young1992cart,le2019senones},
while models with label feedback such as RNN-T and encoder-decoder
%we don't need to have the context encoded in the label,
already cover the (left) context.
Hybrid NN-HMMs also usually split the phonemes % context-dependent phonemes
into begin-middle-end states.
%(usually 3 states) % -- except for silence or some noise tokens, which have a single state).
As we perform the experiments with an encoder-decoder-attention model
which is label-synchronous and auto-regressive,
we also have the end-of-sentence (EOS) token in our vocabulary.
In case of a time-synchronous model,
we possibly would add a blank label
(like for CTC or RNN-T),
or maybe repetition symbols
(as in the Auto Segmentation Criterion (ASG) \cite{collobert2016wav2letter}),
or a silence label (for hybrid NN-HMMs).
It also might make sense to add noise or unknown labels
(e.g.~as in \cite{irie19:phoneme}).

The label units which we are going to compare are:
\begin{itemize}
\item Phonemes
  \begin{itemize}
  \item Single phonemes % (monophones; without context)
  \item Single phonemes + end-of-word (EOW) symbol
  \item Single phonemes + word-end-phones (\#)
%  \item Context-dependent phoneme classes (CART clustering \TODO ref) (not doing this here...)
  \item Phoneme-BPE
  \end{itemize}
  Variation:
  %\begin{itemize}
    %\item Extra end-of-word (EOW) symbol (single phon.)
    %\item
     Extra word-disambiguate symbols
    %\item Extra word-end-phone variants (single phon.)
    %E.g: hello $\rightarrow$ hh ax l ow\# % Note: This example is not clear. You could just say that you add all phones again but with a special marker for EOW, so you end up with twice as many labels.
  %\end{itemize}

\item Graphemes (characters)
  \begin{itemize}
  \item Single characters + whitespace % (with space; without context)
%  \item Context-dependent char classes (not doing this here...)
  \item Char-BPE
  \end{itemize}

\item Whole words
\end{itemize}

Phoneme-BPE is simply the application of BPE on phoneme sequences.
We generate the phoneme BPE codes by taking a single pronunciation for each word
(the most likely as defined by the lexicon)
for all the transcriptions.
The same phoneme sequence is also used for training.
The end-of-word (EOW) symbol for phonemes
is similar as white-space character for grapheme-based models.
%which is standard in that case.
Others \cite{sainath2018nolex,irie19:phoneme} have observed
that the EOW symbol can be helpful for phoneme models.
We follow \cite{zhou2021icassp} to evaluate another variant for single phonemes
where for each phoneme label $x$,
we add another label with word-end-phone marker "\#" yielding "$x$\#".
%where each phoneme label $x$ is augmented with a word-end-phone marker "\#",
%i.e. "$x$\#", if it appears at word end.
This will make the vocab.~size twice larger.

%A phoneme-based model (no matter whether these are single phonemes or subwords)
%cannot be used as-is for recognition.
For phoneme-based models,
we need a lexicon for the mapping of phonemes to words.
But then there are cases
where a phoneme sequence can be mapped to multiple possible words
(\emph{homophones}).
E.g.~the word "I" and "eye" both have the same pronunciation
consisting of the phoneme sequence "ay".
To be able to discriminate between "I" and "eye",
usually an external language model on word-level is used.
Alternatively, we can also add special \emph{word-disambiguate symbols}
to the labels (the phoneme inventory),
in such a way that we can always uniquely discriminate between all words.
%These symbols are not real phonemes -- they are just extra symbols,
%just like EOW.
We go through the pronunciation lexicon and collect all phoneme sequences
which can not be uniquely mapped to words.
For all these phoneme sequences, we add special symbols $\$1$ to $\$N$.
For example
"ay $\$8$" $\rightarrow$ "eye",
"ay $\$9$" $\rightarrow$ "I",
\dots,
"r eh d $\$2$" $\rightarrow$ "read",
"r eh d $\$3$" $\rightarrow$ "red".
%"r eh d $\#4$" $\rightarrow$ "redd".
%\begin{itemize}
%  \item \dots
%  \item ay $\#8$ $\rightarrow$ eye
%  \item ay $\#9$ $\rightarrow$ I
%  \item \dots
%  \item r eh d $\#2$ $\rightarrow$ read
%  \item r eh d $\#3$ $\rightarrow$ red
%  \item r eh d $\#4$ $\rightarrow$ redd
%  \item \dots
%\end{itemize}
These word-disambiguate symbols allow for decoding
without an external language model,
and also allows us to use our simplified decoder implementation.
It also might improve the performance as the model
has now the power to discriminate between words.
Note that this scheme of adding these symbols
does not allow for an easy extension of the lexicon
for further homophones after we trained the model.
%
%Phoneme-BPE with word-disambiguate symbols is also related
%to pronunciation-based subword units \cite{xu2019pronsubword}.
%
%All the phoneme variants require the use of a pronunciation lexicon.
%An additional language model (LM) is mandatory unless we have word-disambiguate symbols.

\section{Model}

We use an encoder-decoder-attention model
\cite{chorowski2015attention,chan2016las}.
Our encoder consists of 6 layers of
bidirectional long short-term memory (LSTM) \cite{hochreiter1997lstm} networks
with time-downsampling via max-pooling by factor 6.
Our decoder is a single layer LSTM network with global attention.
We use SpecAugment \cite{park2019specaugment} for on-the-fly data augmentation.
For further details, please refer to our earlier work
\cite{zeyer2018:asr-attention,zeyer2019:trafo-vs-lstm-asr},
which is exactly the same model,
except for the variations of the output label.

\subsection{Training}

%In all cases,
We maximize the log-likelihood % minimize neg ...
%  \sum_{(x_1^T, y_1^N) \in \mathcal{D}}
$\log p(y_1^N|x_1^T)$,
%which is the standard cross entropy loss,
for target sequences $y_1^N$ and input feature sequences $x_1^T$
over the training dataset.
%from the training dataset $\mathcal{D}$.
We always have a single ground-truth target sequence.
In the case of phonemes,
we reduce the lexicon to contain only a single pronunciation per word,
and thus this becomes unique.
This is a simplification, which we only do for training, not for decoding.
%We could also marginalize over all possible pronunciations in training,
%but that would make it much more complicated,
%and this is also not possible to do efficiently
%without approximations
%such as the maximum approximation.
%We train with stochastic gradient descent (SGD)
%and esp.~the Adam optimizer \cite{kingma2014adam}.
%We use Adam \cite{kingma2014adam} for optimization.
%We do pretraining by starting with a two layer encoder and smaller dimensions,
%and then we grow the encoder in width (dimensions) and depth (number of layers)
%\cite{zeyer2018:attanalysis}.
%Our hyper parameters and training details all follow exactly our earlier work
%\cite{zeyer2018:asr-attention,zeyer2019:trafo-vs-lstm-asr}.

\subsection{Decoding}

Our \emph{simplified decoder} performs  beam search
over the labels
with a fixed beam size (e.g.~12 hypotheses)
without any restrictions (i.e.~it allows any possible label sequence).
Our simplified decoder allows for log-linear combination
with a LM on the same label-level (e.g.~phone-BPE)
but not with word-level LM.
\emph{After} we find a label sequence with this simplified beam search,
we map it to words.
In case of BPE, we first do BPE merging.
In case of phonemes with word-disambiguate symbols,
we try to lookup the corresponding word
(which is unique because of the word-disambiguate symbols),
or replace it by some UNK symbol if not found.
That way, we eventually end up with a sequence of words.

Our \emph{advanced decoder} performs prefix-tree search based on a lexicon. % converted from the official one.
This lexicon defines the mapping between words and corresponding phoneme or grapheme label sequences.
The resulting lexical tree restricts the search to possible label sequences from the lexicon.
It also allows log-linear combination with a word-level LM.
The LM score is applied to a hypothesized path whenever it reaches a word-end.
Optionally, LM lookahead can also be applied to incorporate the LM score into the tree for a more robust search.
The standard beam pruning using a fixed beam size is applied at each search step.
Finally, the decoded best path directly gives the recognized word sequence.

\section{Experiments}

We use RETURNN \cite{zeyer2018:returnn}. %as the training framework.
%which builds upon TensorFlow \cite{tensorflow2015}.
The advanced decoder is implemented as part of RASR \cite{rybach2011:rasr},
while the simplified decoder is implemented within RETURNN. %in pure TensorFlow
All our config files and code to reproduce these experiments
can be found online\footnote{\tiny\url{https://github.com/rwth-i6/returnn-experiments/tree/master/2020-phone-bpe-attention}}.

\subsection{Switchboard 300h}
% total number of utterances: 249K

We perform experiments on the Switchboard 300h English telephone speech
corpus \cite{godfrey1992switchboard}.
We use Hub5'00 as a development set,
which consists of both Switchboard (SWB) and CallHome (CH),
and Hub5'01 as a test set.

During training, because of GPU memory constraints,
we filter out too long sequences
based on a threshold on the target sequence length.
The length of the target sequences varies depending on the labels unit,
e.g.~a sequence of phonemes can be much longer than with grapheme-BPEs.
Thus, to have a fair comparison, we set the maximum allowed
target sequence length for all label units individually
such that we drop the same amount
of training utterances ($0.35\%$ are dropped).

We collect our experiments with our simplified decoder without LM
in \Cref{tab:swb:simple}.
The simplified decoder can only produce reasonable results
if the label units allow for word disambiguations,
such as in the case of graphemes,
but also for phonemes with added word-disambiguate symbols.
We find that BPE subwords perform much better than single units
and also better than whole words,
both for phonemes and graphemes.
In later experiments with the advanced decoder,
where the search is restricted to valid sequences from the lexicon,
we will see that single phoneme units perform much better.
Moreover, word-end-phones perform better than EOW symbol,
which is also different with the advanced decoder.
For grapheme-BPE, BPE-534 model seems to perform best
whereas for phoneme-BPE, both BPE-592 and BPE-1k models are comparable.
%Note that BPE-500 results in 592 phoneme classes, or 534 grapheme classes.
Note that the num.~of labels has an impact on the training time,
due to the softmax output layer, and different output sequence lengths.
Single (EOW) phoneme training is 10-20\% relatively slower than BPEs or whole words.
%Single characters are much slower whereas
%words are comparable with BPEs (depends on vocab size).

\begin{table}[t]
  \caption{Comparing \textbf{phoneme}, \textbf{grapheme} and \textbf{whole word} models on Switchboard 300h. %(using 245K utterances).
  Using simplified decoding with beam size 12 without language model nor lexicon.
  All phoneme models have \emph{word-disambiguate symbols}.
  Single phoneme is with end-of-word (EOW) symbol or word-end-phone symbol (\#).
  Single grapheme is with whitespace. %, which is like a EOW symbol.
  }
  \label{tab:swb:simple}
  \centering
  \setlength{\tabcolsep}{0.3em}
  \def\arraystretch{0.95}
  % For every single line in the table with some WER numbers,
  % make a comment (before the line) which setup (config name) that is exactly, and which epoch.
  \resizebox{\columnwidth}{!}{%
  \begin{tabular}{|c|c|c|S[table-format=2.1]|S[table-format=2.1]|S[table-format=2.1]|S[table-format=2.1]|}
    \hline
    \multicolumn{3}{|c|}{Labels} &\multicolumn{4}{c|}{WER[\%]} \\ \cline{1-7}
    \multirow{2}{*}{Unit} & \multirow{2}{*}{Type} & \multirow{2}{*}{\#Num}
    & \multicolumn{3}{c|}{Hub5'00} & \multicolumn{1}{c|}{Hub5'01} \\ \cline{4-7}
    & & & SWB & CH & $\Sigma$ & $\Sigma$ \\
    \hline
    % config path : /u/zeineldeen/setups/switchboard/2020-10-10--att-phon-paper/config-train/base2.conv2l.specaug4.phone_orth_eow.disamb.msl228.config
    % epoch : 190 , #num : 62
    \multirow{8}{*}{Phoneme} & Single (EOW) & 62 & 12.6 & 30.7 & 21.7 & 18.2 \\
    \cline{2-7}
    % /u/zeineldeen/setups/switchboard/2020-10-10--att-phon-paper/config-train/base2.conv2l.specaug4.phone_orth.end_phon.disamb.end_phon.msl180.fix.config
    % epoch: 200, #num: 118
                             & \multicolumn{1}{l|}{Single (\#)} & 118 & 11.6 & 25.7 & 18.7 & 16.7 \\
    \cline{2-7}
    % config path : /u/zeineldeen/setups/switchboard/2020-10-10--att-phon-paper/config-train/phone-uni-bpe50.base2.conv2l.specaug4a.disamb.mql130.config
    % epoch : 199, #num : 151
    & \multirow{6}{*}{BPE} & 151 & 10.9 & 23.9 & 17.4 & 16.6 \\
    \cline{3-7}
    % config path : /u/zeineldeen/setups/switchboard/2020-10-10--att-phon-paper/config-train/phone-uni-bpe100.base2.conv2l.specaug4a.disamb.mql143.config
    % epoch : 160, #num : 201
    &  & 201 & 10.6 & 22.8 & 16.7 & 15.9 \\
    \cline{3-7}
    % config path : /u/tng/setups/switchboard/2019-07-29--att-bpe1k/config-train/phone-uni-bpe500.base2.conv2l.specaug4a.disamb.mql125.config
    % epoch : 182, #num : 592, bpe 500
    &  & 592 & 9.7 & 20.8 & \B 15.3 & 14.9 \\
    \cline{3-7}
    % config path : /u/tng/setups/switchboard/2019-07-29--att-bpe1k/config-train/phone-uni-bpe1k.base2.conv2l.specaug4a.disamb.mql110.config
    % epoch : 200, #num : 1091
    & & 1k & 10.0 & 20.4 & \B 15.2 & 15.1 \\
    \cline{3-7}
    % config path : /u/tng/setups/switchboard/2019-07-29--att-bpe1k/config-train/phone-uni-bpe2k.base2.conv2l.specaug4a.disamb.mql96.config
    % epoch : 200, #num : 2086
    & & 2k & 10.3 & 21.4 & 15.9 & 15.4 \\
    \cline{3-7}
    % config path : /u/tng/setups/switchboard/2019-07-29--att-bpe1k/config-train/phone-uni-bpe5k.base2.conv2l.specaug4a.disamb.mql87.config
    % epoch : 191, #num : 5033
    & & 5k & 10.5 & 21.6 & 16.0 & 15.5 \\
    \hline \hline
    % config path : /u/tng/setups/switchboard/2019-07-29--att-bpe1k/config-train/base2.conv2l.specaug4.char.mql250.config
    % epoch : 160, #num : 35
    \multirow{9}{*}{Grapheme} & Single & 35 & 13.3 & 32.8 & 23.1 & 18.6 \\
    \cline{2-7}
    % config path : /u/tng/setups/switchboard/2019-07-29--att-bpe1k/config-train/base2.conv2l.specaug4.bpe50.mql154.config
    % epoch : 197, #num : 126, bpe 50
    & \multirow{8}{*}{BPE} & 126 & 10.3 & 22.9  & 16.6  & 15.8    \\
    \cline{3-7}
     % config path : /u/tng/setups/switchboard/2019-07-29--att-bpe1k/config-train/base2.conv2l.specaug4.bpe100.mql134.config
    % epoch : 190, #num : 176, bpe 100
    &   & 176 & 9.7 & 21.0 & 15.3 & 14.7   \\
    \cline{3-7}
    % config path : /u/tng/setups/switchboard/2019-07-29--att-bpe1k/config-train/base2.conv2l.specaug4.bpe500.mql88.config
    % epoch : 199, #num : 534, bpe 500
    & & 534 & 10.0 & 20.6  & \B 15.3 & 14.8 \\
    \cline{3-7}
    % config path : /u/tng/setups/switchboard/2019-07-29--att-bpe1k/config-train/base2.conv2l.specaug4.bpe-1k.config
    % epoch : 186, #num : 1030
    % using all data: Hub5'00 --> 15.4, Hub5'01 --> 15
    & & 1k & 10.3 & 21.1 & 15.7 & 15.5 \\
    \cline{3-7}
    % config path : /u/tng/setups/switchboard/2019-07-29--att-bpe1k/config-train/base2.conv2l.specaug4.bpe-2k.mql67.config
    % epoch : 195, #num : 2026
    & & 2k & 10.0 & 21.4 & 15.7 & 14.7  \\
    \cline{3-7}
    % config path : /u/tng/setups/switchboard/2019-07-29--att-bpe1k/config-train/base2.conv2l.specaug4.bpe-5k.mql60.config
    % epoch : 179, #num : 4980
    & & 5k & 10.4 & 21.8  & 16.1 & 15.5  \\
    \cline{3-7}
    % config path : /u/tng/setups/switchboard/2019-07-29--att-bpe1k/config-train/base2.conv2l.specaug4.bpe-10k.mql57.config
    % epoch : 173, #num : 9795
    &   & 10k & 10.9 & 22.7 & 16.8 &15.9     \\
    \cline{3-7}
    % config path : /u/tng/setups/switchboard/2019-07-29--att-bpe1k/config-train/base2.conv2l.specaug4.bpe-20k.mql56.config
    % epoch : 162, #num : 18611
    &   & 20k & 11.4 & 22.9 & 17.2 & 16.6     \\
    \hline \hline
    % config path : /u/zeineldeen/setups/switchboard/2020-10-10--att-phon-paper/config-train/base2.conv2l.specaug4.words_orth.mql55.config
    % epoch : 173
    % Note: this is worst than using 75 (was 18.1 on dev). also, with no msl,
    % we can get here 17.4 on dev
    Words & Single & 30k & 12.5 & 25.0 & 18.8 & 18.0 \\
    \hline
  \end{tabular}
  }
\end{table}

We compare different decoding variants in \Cref{tab:swb:decoder}.
%In all further experiments,
%we use
Our advanced decoder % allows to use a word-level LM and
restricts the search to only label sequences
which occur in the lexicon,
including only the BPE-splits seen during training,
in contrast to the simplified decoder,
which does not have this restriction.
%which can be an advantage in certain cases
%but also a disadvantage.
We see that in the case of single phone or grapheme labels,
i.e.~where we have a weaker model,
the restriction on the lexicon by the advanced decoder is significantly helpful
(consistent with \cite{zweig2017allneural})
while it is hurtful for the BPE variants,
esp.~in the case of phoneme-BPE.
%Also in the case of graphemes,
%we could restrict the search to words in a vocabulary,
%which might improve the performance in certain cases \cite{zweig2017allneural}.
% Note: This comment about BPE, we don't do, and also don't have a ref to anyone doing this, so let's skip for now (space issues).
%In the case of BPE (either phoneme or grapheme),
%we can also restrict the search to BPE splits seen during training,
%which might reduce unexpected behavior of the decoder,
%but which might also decrease the performance.
%
%We see that the restriction on the lexicon and
%esp.~on the single unique greedy BPE-split in the advanced beam search decoder
%can be hurtful in some cases,
%e.g.~for phone-BPE,
%where the effect is more hurtful than expected.
We also see that the word-level LM improves
the performance in all cases (as expected).

% NOTE: The best results used beam size of 12.
%       here we don't get improvement with larger beam sizes and thus their
%       results are dropped from this table. this is probably due to length bias
\begin{table}[t]
  \caption{\textbf{Decoding comparison} on Switchboard 300h.
  All phoneme models here have word-disambiguate symbols.
  Single phoneme is with EOW symbol.
  Single grapheme is with whitespace. %, which is like a EOW symbol.
  All with beam size 12 and the optional LSTM LM is on word-level.
  The advanced decoder is also restricted on the lexicon
  and the unique greedy BPE-split.
  }
  \label{tab:swb:decoder}
  \centering
  \setlength{\tabcolsep}{0.2em}
  % For every single line in the table with some WER numbers,
  % make a comment (before the line) which setup (config name) that is exactly, and which epoch.
  \resizebox{\columnwidth}{!}{%
  \begin{tabular}{|c|c|c|c|S[table-format=2.1]|S[table-format=2.1]|S[table-format=2.1]|S[table-format=2.1]|}
    \hline
    \multicolumn{2}{|c|}{Labels} & \multicolumn{2}{c|}{Decoder} & \multicolumn{4}{c|}{WER[\%]} \\
    \cline{1-8}
    \multirow{2}{*}{Unit} & \multirow{2}{*}{Type} & \multirow{2}{*}{LM} & \multirow{2}{*}{Impl.} & \multicolumn{3}{c|}{Hub5'00} & \multicolumn{1}{c|}{Hub5'01} \\ \cline{5-8}
    && & & SWB & CH & $\Sigma$ & $\Sigma$ \\
    \hline
    % models and epoch are the same as in Table 1
    \multirow{6}{*}{Phon.} & \multirow{2}{*}{Single} & None & Simplified & 12.6 & 30.7 & 21.7 & 18.2 \\
    \cline{4-8}
    & \multirow{2}{*}{(EOW)} & & \multirow{2}{*}{Advanced} & 9.9 & 21.2 & 15.6 & 15.2 \\ \cline{3-3} \cline{5-8}
    && LSTM & & 9.0 & 20.6 & \B 14.8 & 14.1 \\ \cline{3-4} \cline{5-8}
    % && & 32 & & 9.3 & 20.5 & 14.9 \\ \cline{4-4} \cline{6-8}
    % && & 64 & & 9.3 & 20.5 & 14.9 \\
    \cline{2-8}

    & \multirow{3}{*}{BPE-592} & None & Simplified & 9.7 & 20.8 & 15.3 & 14.9 \\
    \cline{4-8}
    && & \multirow{2}{*}{Advanced} & 10.5 & 22.3 & 16.4 & 16.1 \\ \cline{3-3} \cline{5-8}
    && LSTM & & 9.5 & 21.6 & 15.6 & 14.8 \\ \cline{3-4} \cline{5-8} \hline \hline
    % && & 32 & & 9.9 & 21.9 & 15.9 \\ \cline{4-4} \cline{6-8}
    % && & 64 & & 10.1 & 22.6 & 16.3 \\ \hline

    \multirow{6}{*}{Graph.} & \multirow{3}{*}{Single} & None & Simplified & 13.3 & 32.8 & 23.1 & 18.6 \\
    \cline{4-8}
    && & \multirow{2}{*}{Advanced} & 10.4 & 21.9 & 16.2 & 15.4 \\ \cline{3-3} \cline{5-8}
    && LSTM & & 9.3 & 21.1 & 15.3 & 14.2 \\ \cline{3-4} \cline{5-8}
    % && & 32 & & 9.4 & 21.3 & 15.4 \\ \cline{4-4} \cline{6-8}
    % && & 64 & & 9.4 & 21.4 & 15.4 \\
    \cline{2-7}

    & \multirow{3}{*}{BPE-534} & None & Simplified & 10.0 & 20.6 & 15.3 & 14.8 \\
    \cline{4-8}
    && & \multirow{2}{*}{Advanced} & 10.1 & 20.5 & 15.3 & 15.0 \\ \cline{3-3} \cline{5-8}
    && LSTM & & 8.9 & 20.2 & \B 14.6 & 13.9 \\ \cline{3-4} \cline{5-8}
    % && & 32 & & 8.9 & 20.3 & 14.7 \\ \cline{4-4} \cline{6-8}
    % && & 64 & & 9.3 & 20.7 & 15.0  \\
    \hline
  \end{tabular}
  }
\end{table}

We study the effect of the word-disambiguate symbols for phoneme-based models
in \Cref{tab:swb:phone-disambig}.
We find that the word-disambiguate symbols seem to be hurtful.
We are still careful in drawing conclusions from this,
as this might be due to the specific variant of how we added such symbols.

For single phoneme labels,
we also study the effect of having an end-of-word (EOW) symbol or not,
or using the word-end-phones variant.
All the results are with the advanced decoder,
i.e.~restricted only to valid entries from the lexicon,
and with LM.
We show the results in \Cref{tab:swb:compare-eow-and-word-end-phon}. % \Cref{tab:swb:single-phone-eow}.
Consistent with \cite{irie19:phoneme}, having EOW helps.
In this setting, we observe that word-end-phones are worse,
in contrast to the simplified decoder without restriction from the lexicon.

\begin{table}[t]
  \caption{Studying \textbf{word-disambiguate symbols}
  on Switchboard 300h by comparing different \textbf{phoneme} variants.
  All with word-level LSTM LM, and the advanced decoder.
  }
  \label{tab:swb:phone-disambig}
  \centering
  \setlength{\tabcolsep}{0.3em}
  % For every single line in the table with some WER numbers,
  % make a comment (before the line) which setup (config name) that is exactly, and which epoch.
  \begin{tabular}{|c|c|c|c|S[table-format=2.1]|S[table-format=2.1]|S[table-format=2.1]|}
    \hline
    \multicolumn{3}{|c|}{Phoneme Labels} & \multirow{2}{*}{Beam} & \multicolumn{3}{c|}{WER[\%]} \\
    \cline{1-3} \cline{5-7}
    \multirow{2}{*}{Type} & \multirow{2}{*}{\#Num} & \multirow{2}{*}{Disamb.} & \multirow{2}{*}{size} & \multicolumn{3}{c|}{Hub5'00} \\
    \cline{5-7}
    & & & & SWB & CH & $\Sigma$ \\
    \hline
    % same model as Table 2
    \multirow{2}{*}{Single} & \multirow{2}{*}{62} & \multirow{2}{*}{Yes} & 12 & 9.0 & 20.6 & 14.8 \\ \cline{4-7}
    \multirow{2}{*}{(EOW)}  &    &   & \multirow{2}{*}{32} & 9.3 & 20.5 & 14.9 \\ \cline{2-3} \cline{5-7}
    % /u/zeineldeen/setups/switchboard/2020-10-10--att-phon-paper/config-train/base2.conv2l.specaug4.phone_orth_eow.wo_disamb.msl228.config
    % epoch: 184, lm-scale: 0.2, beam-size: 32
    % NOTE: with beam size 12: 8.8 & 21 & 14.9
    %       with beam size 32/64: 8.7 & 20.1 & 14.4

        & 48 & {No} & & \B 8.7 & \B 20.1 & \B 14.4 \\ \hline \hline

    % same model as Table 2
    \multirow{3}{*}{BPE} & \multirow{2}{*}{592} & \multirow{2}{*}{Yes} & 12 & 9.5 & 21.6 & 15.6 \\ \cline{4-7}
    & & & \multirow{2}{*}{64} & 10.1 & 22.6 & 16.3 \\ \cline{3-3} \cline{5-7}
    \cline{2-2}
    % /u/zeineldeen/setups/switchboard/2020-01-21--att-phon/config-train/base2.conv2l.specaug4a.phone-uni-bpe500.max_seq_len100.config
    % epoch: 197
           & 588 & No & & 8.7 & 20.7 & 14.7 \\ \hline
  \end{tabular}
\end{table}

%\begin{table}[t]
%\caption{On Switchboard 300h, WER on Hub 5'00.
%Comparing \textbf{end-of-word (EOW)} for single phonemes,
%without word-disambiguate symbols,
%with and without LM.
%All with beam size 32.
%In case of no LM, when there are multiple words
%corresponding to the same phone sequence,
%the decoder will just pick the first (alphabetically).
%}
%\label{tab:swb:single-phone-eow}
%\centering
%\setlength{\tabcolsep}{0.3em}
%\begin{tabular}{|c|c|c|}
%\hline
%%Additionally, for the single-phon EOW exps:
%% - without LM, noEOW does not work at all (220% WER) even lexicon
%
%% /u/zeineldeen/setups/switchboard/2020-10-10--att-phon-paper/config-train/base2.conv2l.specaug4.phone_orth_eow.wo_disamb.msl228.config
%% epoch: 160
%
%LM & EOW & WER[\%] \\ \hline
%\multirow{2}{*}{No} & No & $>$100\phantom{.0} \\ \cline{2-3}
%                    & Yes & \phantom{$>$1}36.5 \\ \cline{2-3} \cline{1-1}
%\multirow{2}{*}{Yes} & No & \phantom{$>$1}14.8 \\ \cline{2-3}
%                     & Yes & \phantom{$>$1}\textbf{14.4} \\ \cline{2-3}
%\hline
%\end{tabular}
%\end{table}

\begin{table}[t]
\caption{
Comparing \textbf{end-of-word (EOW)}
and \textbf{word-end-phones} variant for single phonemes.
%with \textbf{word disambiguate symbols} or without.
WER is on Hub5'00.
All results are with word-level LM and advanced decoder, beam size 32,
without word disambiguate symbols.
}
\label{tab:swb:compare-eow-and-word-end-phon}
\centering
\begin{tabular}{|c|c|c|}
\hline
Variant & \#Num & WER[\%] \\ \hline
No EOW & 47 & 14.8 \\ \hline
EOW   & 48 & \B 14.4 \\ \hline
Word-end-phones (\#) & 90 & 15.5 \\ \hline
\end{tabular}
%\begin{tabular}{|c|c|c|}
%\hline
%Disamb. & Variant & WER[\%] \\ \hline
%\multirow{2}{*}{Yes} & EOW & 14.9 \\ \cline{2-3}
% & Word-end-phones & 17.6 \\ \hline
%
%\multirow{3}{*}{No} & No EOW & 14.8 \\ \cline{2-3}
% & EOW   &  14.4 \\ \cline{2-3}
% & Word-end-phones & 15.5 \\
%\hline
%\end{tabular}
\end{table}

Finally, we compare our results to other results from the literature
in \Cref{tab:swb:literature}.
We observe that many other works train for much longer,
and that seems to lead to yet better results.
Our phoneme-based models perform slightly better
than our final grapheme-based models,
although they are very close.
%We also get better result with phoneme-BPE when using all training sequences. % NOTE: Of course this is the case, no need to mention.
%With other label units, this improvement was not observed. % NOTE: Confusing, wrong, misleading. Just a matter of bad tuning.

\begin{table}[t]
  \caption{Comparing results from the \textbf{literature}
  on Switchboard 300h.
  One big difference in varying results
  is the different amount of training time (number of epochs).
  $^*$no train seq.~len.~filter.
  }
  \label{tab:swb:literature}
  \centering
  \resizebox{\columnwidth}{!}{%
%\begin{adjustbox}{width=1.\width,center}
%  \setlength{\tabcolsep}{0.3em}
\setlength{\tabcolsep}{0.1em}
  % For every single line in the table with some WER numbers,
  % make a comment (before the line) which setup (config name) that is exactly, and which epoch.
  % Note: We can infer Hub500 total from individual SWB/CH:
  % Num words SWB: 21399, CH: 21594.
  % So factor 0.497732189 for SWB, factor 0.502267811 for CH.
  \begin{tabular}{|c|c|c|c|c|c|S[table-format=2.1]|S[table-format=2.1]|S[table-format=2.1]|S[table-format=2.1]|S[table-format=2.1]|c|c|}
    \hline
    \multirow{3}{*}{Work} & \multicolumn{3}{c|}{Label} & \multirow{3}{*}{\#Ep}
    & \multirow{3}{*}{LM} & \multicolumn{5}{c|}{WER[\%]} \\
    \cline {2-4} \cline{7-11}

    &  \multirow{2}{*}{Unit} & \multirow{2}{*}{Type} & \multirow{2}{*}{\#Num}
    & & & \multicolumn{3}{c|}{Hub5$^{00}\!$} & \multicolumn{1}{c|}{Hub5$^{01}\!$} & RT$^{03}\!$ \\
    \cline{7-11}

    &&& && & SWB & CH & $\Sigma$ & $\Sigma$ & $\Sigma$ \\
    \hline
    \cite{zeineldeen:icassp20} & Phon. & CART & 4.5k & \phantom{0}13 & Yes & 9.6 & 18.5 & 14.0 & 14.1 & \\
    \hline

    % Zeyer Trafo-vs-LSTM, with LM: BPE & 1k & \phantom{0}33 & 9.3 & 20.3 & 14.9 & 14.2 &
    % without LM: 10.1 & 20.6 & 15.4 & 14.7
    \cite{zeyer2019:trafo-vs-lstm-asr} & Graph. & BPE & 1k & \phantom{0}33 & No & 10.1 & 20.6 & 15.4 & 14.7 & \\
    \cline{6-11}
     &  &  &  & & Yes & 9.3 & 20.3 & 14.9 & 14.2 & \\
    \cline{1-1}\cline{4-11}

    % https://arxiv.org/abs/1910.13296. results are without LM.
    \cite{nguyen2019improving} &  &  & 4k & \phantom{0}50 & No & 8.8 & 17.2 & 13.0 && \\
    \cline{1-1}\cline{4-11}

    % https://github.com/espnet/espnet/blob/master/egs/swbd/asr1/RESULTS.md
    % with LM: BPE & 2k & 100 & 9.0 & 18.1 & 13.6
    \cite{karita2019trafo} &  &  & 2k & 100 & Yes & 9.0 & 18.1 & 13.6 & & \\
    \cline{1-2}\cline{4-11}

    \cite{wang2020phonebpe} & Phon. &  & 500 & 150 & Yes & 7.9 & 16.1 & 12.0 && 14.5 \\
    \cline{1-2}\cline{4-11}

    % single headed att, tuske (https://arxiv.org/pdf/2001.07263.pdf):
    % with LM, and xutt: 6.4 & 12.5 &
    % without LM: 7.6 & 14.6; hub501 swb 8.1, swb2p3 11.0, swb2p4 15.7; rt03 swb 17.8, fsh 10.5
    \cite{tuske2020swbatt} & Graph. &  & 600 & 250 & No & 7.6 & 14.6 & 11.1 && \\
    \cline{6-11}
    &  &  &  &  & Yes & 6.4 & 12.5 & 9.5 && \\
    \cline{1-1}\cline{3-11}

    % specaug: 340k steps, 512 seqs/step, 227047 seqs -> ~340000/(227047/512) = 766 epochs
    % specaug, with LM:  WPM &1k & 760 & 6.8 & 14.1 & &&
    % specaug, without LM: 7.2 & 14.6
    \cite{park2019specaugment} &  & WPM &1k & 760 & No & 7.2 & 14.6 & 10.9 && \\
    \cline{6-11}
     &  &  & &  & Yes & 6.8 & 14.1 & 10.5 && \\

    \hline
    \hline
   % \multicolumn{1}{|c}{Ours} & \multicolumn{10}{c|}{} \\ \hline
    %\multirow{4}{*}{Ours}
    Ours

    % config path : /u/tng/setups/switchboard/2019-07-29--att-bpe1k/config-train/base2.conv2l.specaug4.bpe500.mql88.config
    % epoch : 199, #num : 534, bpe 500
    & Graph & BPE & 534 & \phantom{0}33 &   No & 10.1 & 20.5 & 15.3 & 15.0 & 18.0 \\ \cline{6-11}
    &        &       &     &  & Yes & 8.9 & 20.2 & 14.6 & 13.9 & 16.9 \\
    \cline{2-4}\cline{7-11}

    % /u/zeineldeen/setups/switchboard/2020-10-10--att-phon-paper/config-train/base2.conv2l.specaug4.phone_orth_eow.wo_disamb.msl228.config
    % epoch: 184
    & Phon. & Single & 47 &  &  & \B 8.7 & 20.1 & 14.4 & \B 13.8 & 16.7 \\
    \cline{3-4} \cline{7-11}

    % /u/zeineldeen/setups/switchboard/2020-01-21--att-phon/config-train/base2.conv2l.specaug4a.phone-uni-bpe500.max_seq_len100.config
    % epoch: 197
    &       & BPE & 588 &  &  & \B 8.7 & 20.7 & 14.7 & \B 13.8 & 17.5 \\
    \cline{1-1} \cline{7-11}

    Ours{$^*$} & & & & & & \B 8.7 & \B 19.7 & \B 14.2 & \B 13.8 & \B 16.2 \\
    \hline
  \end{tabular}
  %\end{adjustbox}
  }
\end{table}

\subsection{LibriSpeech 960h}

Furthermore, we conduct experiments on the LibriSpeech 960h corpus
\cite{Panayotov2015librispeech}.
All models are trained for 13 epochs.
%
%We evaluate the models on the dev and test datasets, namely dev-other,
%dev-clean, test-other, and test-clean.
%
Our results are in \Cref{tab:libri:compare}.
We observe that grapheme and phoneme BPE outperform single phonemes,
in contrast to our Switchboard results.
This might be due to the $0.7\%$ OOV rate. %and unknown labels.
We also observed that specifically
the single phoneme model overfits much more than all other experiments.
%Thus, this might bias the effect of the external LM used.
Grapheme BPE-10k reports the best WER on all data sets.

\begin{table}[t!]
\caption{Comparing models with different label units on LibriSpeech 960h.
All phoneme models are without word-disambiguate symbols.
All with word-level LSTM LM, and the advanced decoder.
}
\label{tab:libri:compare}
\centering

% For every single line in the table with some WER numbers,
% make a comment (before the line) which setup (config name) that is exactly, and which epoch.
\resizebox{\columnwidth}{!}{%
\setlength{\tabcolsep}{0.4em}
  \def\arraystretch{0.95}
\begin{tabular}{|c|c|c|S[table-format=1.2]|S[table-format=2.2]|S[table-format=1.2]|S[table-format=2.2]|}
  \hline
  \multicolumn{3}{|c|}{Labels} & \multicolumn{4}{c|}{WER[\%]} \\ \cline{1-7}
  \multirow{2}{*}{Unit} & \multirow{2}{*}{Type} & \multirow{2}{*}{\#Num} &
  \multicolumn{2}{c|}{dev} & \multicolumn{2}{c|}{test} \\ \cline{4-7}
  & & & \text{clean} & \text{other} & \text{clean} & \text{other} \\
  \hline
  \multirow{4}{*}{Phoneme} & Single (\#) & 141 & 5.14 & 12.83 & 5.71 & 13.79 \\ \cline{2-7}
  & \multirow{3}{*}{BPE} & 5k & 3.42 & 8.91 & 3.91 & 9.74 \\ \cline{3-7}
  &                      & 10k & 3.28 & 8.82 & 3.88 & 9.31 \\ \cline{3-7}
  &                      & 20k & 3.50 & 9.06 & 3.86 & 10.56 \\
  \hline \hline
  \multirow{3}{*}{Grapheme} & \multirow{3}{*}{BPE} & 5k & 3.35 & 8.41 & 3.84
  & 9.48 \\ \cline{3-7}
  &                         & 10k & \B 3.15 & \B 8.32 & \B 3.59 & \B 9.14  \\ \cline{3-7}
  &                         & 20k & 3.46 & 9.35 & 3.93 & 10.08 \\
  \hline
\end{tabular}
}
\end{table}

\section{Conclusions}

We compared phoneme-based labels vs.~grapheme-based labels
for attention-based encoder-decoder models
and found their performance to be similar
-- the phoneme-based models are slightly better
on the smaller Switchboard corpora,
but worse on the LibriSpeech corpora.
We also compared single units vs.~subword (BPE) units vs.~whole words,
and found that subword units are best, both for phonemes and graphemes.
%While this was already well-known for grapheme-based models,
%this is a new observation for phoneme-based models.  % NOTE: Confusing sentence...
For single units,
the restriction to valid sequences via lexicon in search
%(via our advanced decoder)
is crucial for good performance.

\section{Acknowledgements}

% Note to authors: Authors should not use logos in acknowledgement section; rather authors should acknowledge corporations by naming them only.

\begin{spacing}{0.8}
%\footnotesize
%\begin{footnotesize}
%\setstretch{0.8}
This work has received funding
from the European Research Council (ERC)
under the European Union’s Horizon 2020 research
and innovation programme
(grant agreement No 694537, project ”SEQCLAS”)
and from a Google Focused Award.
The work reflects only the authors’ views and none of
the funding parties is responsible for any use that may be
made of the information it contains.
%\end{footnotesize}
%\normalsize
\end{spacing}
\setstretch{0.2}

% \vfill
% \pagebreak

%\def\baselinestretch{0.8}
%\SetTracking{encoding=*}{-15}\lsstyle  % still somewhat ok
%\SetTracking{encoding=*}{-85}\lsstyle

%\renewcommand{\baselinestretch}{0.1}\normalsize
\newcommand{\myfont}{\fontsize{8.6}{10.1}\selectfont}
% http://tex.stackexchange.com/questions/93859/condense-the-space-between-bibliographic-entries
\let\OLDthebibliography\thebibliography
\renewcommand\thebibliography[1]{
  \OLDthebibliography{#1}
  \setlength{\parskip}{0.5pt}
  \setlength{\itemsep}{0.5pt}
  \setstretch{0.8}
  \myfont
}

% References should be produced using the bibtex program from suitable
% BiBTeX files (here: strings, refs, manuals). The IEEEbib.bst bibliography
% style file from IEEE produces unsorted bibliography list.
% -------------------------------------------------------------------------

\bibliographystyle{IEEEbib}

\bibliography{phone-paper}

\begin{thebibliography}{10}

\bibitem{bahdanau2015align}
D. Bahdanau, K. Cho, and Y. Bengio,
\newblock ``Neural machine translation by jointly learning to align and
  translate,''
\newblock in {\em ICLR}, 2015.

\bibitem{luong2015attentionmt}
M.-T. Luong, H. Pham, and C.~D. Manning,
\newblock ``Effective approaches to attention-based neural machine
  translation,'' Preprint arXiv:1508.04025, 2015.

\bibitem{zeyer2018:asr-attention}
A. Zeyer, K. Irie, R. Schlüter, and H. Ney,
\newblock ``Improved training of end-to-end attention models for speech
  recognition,''
\newblock in {\em Interspeech}, 2018.

\bibitem{park2019specaugment}
D.~S. Park, W. Chan, Y. Zhang, C.-C. Chiu, B. Zoph, E.~D. Cubuk, and Q.~V. Le,
\newblock ``{SpecAugment}: A simple data augmentation method for automatic
  speech recognition,''
\newblock in {\em Interspeech}, 2019.

\bibitem{zeyer2019:trafo-vs-lstm-asr}
A. Zeyer, P. Bahar, K. Irie, R. Schlüter, and H. Ney,
\newblock ``A comparison of {Transformer} and {LSTM} encoder decoder models for
  {ASR},''
\newblock in {\em ASRU}, 2019.

\bibitem{tuske2020swbatt}
Z. T{\"u}ske, G. Saon, K. Audhkhasi, and B. Kingsbury,
\newblock ``Single headed attention based sequence-to-sequence model for
  state-of-the-art results on switchboard-300,'' Preprint arXiv:2001.07263,
  2020.

\bibitem{graves2006connectionist}
A. Graves, S. Fern{\'a}ndez, F. Gomez, and J. Schmidhuber,
\newblock ``Connectionist temporal classification: labelling unsegmented
  sequence data with recurrent neural networks,''
\newblock in {\em ICML}, 2006.

\bibitem{miao2015eesen}
Y. Miao, M. Gowayyed, and F. Metze,
\newblock ``{EESEN}: End-to-end speech recognition using deep {RNN} models and
  {WFST}-based decoding,''
\newblock in {\em ASRU}. IEEE, 2015.

\bibitem{amodei2015deepspeech2}
D. Amodei, R. Anubhai, E. Battenberg, C. Case, J. Casper, B. Catanzaro, J.
  Chen, M. Chrzanowski, A. Coates, G. Diamos, et~al.,
\newblock ``Deep speech 2: End-to-end speech recognition in english and
  mandarin,'' Preprint arXiv:1512.02595, 2015.

\bibitem{zweig2017allneural}
G. Zweig, C. Yu, J. Droppo, and A. Stolcke,
\newblock ``Advances in all-neural speech recognition,''
\newblock in {\em ICASSP}, 2017.

\bibitem{graves2012seqtransduction}
A. Graves,
\newblock ``Sequence transduction with recurrent neural networks,'' Preprint
  arXiv:1211.3711, 2012.

\bibitem{zeyer2020:transducer}
A. Zeyer, A. Merboldt, R. Schlüter, and H. Ney,
\newblock ``A new training pipeline for an improved neural transducer,''
\newblock in {\em Interspeech}, 2020.

\bibitem{gage1994bpe}
P. Gage,
\newblock ``A new algorithm for data compression,''
\newblock {\em C Users Journal}, vol. 12, no. 2, 1994.

\bibitem{sennrich2015neuralbpe}
R. Sennrich, B. Haddow, and A. Birch,
\newblock ``Neural machine translation of rare words with subword units,''
  Preprint arXiv:1508.07909, 2015.

\bibitem{schuster2012wordpieces}
M. Schuster and K. Nakajima,
\newblock ``{Japanese} and {korean} voice search,''
\newblock in {\em ICASSP}, 2012.

\bibitem{wu2016googlenmt}
Y. Wu, M. Schuster, Z. Chen, Q.~V. Le, M. Norouzi, W. Macherey, M. Krikun, Y.
  Cao, Q. Gao, K. Macherey, et~al.,
\newblock ``Google's neural machine translation system: Bridging the gap
  between human and machine translation,'' Preprint arXiv:1609.08144, 2016.

\bibitem{kudo2018subword}
T. Kudo,
\newblock ``Subword regularization: Improving neural network translation models
  with multiple subword candidates,''
\newblock in {\em ACL}, 2018.

\bibitem{bostrom2020unigramlm}
K. Bostrom and G. Durrett,
\newblock ``Byte pair encoding is suboptimal for language model pretraining,''
  Preprint arXiv:2004.03720, 2020.

\bibitem{chan2016lsd}
W. Chan, Y. Zhang, Q. Le, and N. Jaitly,
\newblock ``Latent sequence decompositions,'' Preprint arXiv:1610.03035, 2016.

\bibitem{drexler2020subword}
J. {Drexler} and J. {Glass},
\newblock ``Learning a subword inventory jointly with end-to-end automatic
  speech recognition,''
\newblock in {\em ICASSP}, 2020.

\bibitem{xu2019pronsubword}
H. Xu, S. Ding, and S. Watanabe,
\newblock ``Improving end-to-end speech recognition with pronunciation-assisted
  sub-word modeling,''
\newblock in {\em ICASSP}, 2019.

\bibitem{soltau2017wordctc}
H. Soltau, H. Liao, and H. Sak,
\newblock ``Neural speech recognizer: Acoustic-to-word {LSTM} model for large
  vocabulary speech recognition,''
\newblock in {\em Interspeech}, 2017.

\bibitem{audhkhasi2017wordctc}
K. Audhkhasi, B. Ramabhadran, G. Saon, M. Picheny, and D. Nahamoo,
\newblock ``Direct acoustics-to-word models for english conversational speech
  recognition,''
\newblock in {\em Interspeech}, 2017.

\bibitem{palaskar2018word}
S. Palaskar and F. Metze,
\newblock ``Acoustic-to-word recognition with sequence-to-sequence models,''
\newblock in {\em 2018 IEEE Spoken Language Technology Workshop (SLT)}. IEEE,
  2018.

\bibitem{chen2018modular}
Z. Chen, Q. Liu, H. Li, and K. Yu,
\newblock ``On modular training of neural acoustics-to-word model for
  {LVCSR},''
\newblock in {\em ICASSP}, 2018.

\bibitem{audhkhasi2018building}
K. Audhkhasi, B. Kingsbury, B. Ramabhadran, G. Saon, and M. Picheny,
\newblock ``Building competitive direct acoustics-to-word models for english
  conversational speech recognition,''
\newblock in {\em ICASSP}, 2018.

\bibitem{das2019wordctc}
A. Das, J. Li, G. Ye, R. Zhao, and Y. Gong,
\newblock ``Advancing acoustic-to-word {CTC} model with attention and
  mixed-units,''
\newblock {\em IEEE/ACM Transactions on Audio, Speech, and Language
  Processing}, vol. 27, no. 12, 2019.

\bibitem{bruguier2019phoebe}
A. Bruguier, R. Prabhavalkar, G. Pundak, and T.~N. Sainath,
\newblock ``Phoebe: Pronunciation-aware contextualization for end-to-end speech
  recognition,''
\newblock in {\em ICASSP}, 2019.

\bibitem{bourlard1994hybrid}
H. Bourlard and N. Morgan,
\newblock {\em Connectionist speech recognition: a hybrid approach},
\newblock Springer, 1994.

\bibitem{robinson1994rnnhmm}
A.~J. Robinson,
\newblock ``An application of recurrent nets to phone probability estimation,''
\newblock {\em Neural Networks, IEEE Transactions on}, vol. 5, no. 2, 1994.

\bibitem{zeyer17:lstm}
A. Zeyer, P. Doetsch, P. Voigtlaender, R. Schlüter, and H. Ney,
\newblock ``A comprehensive study of deep bidirectional {LSTM} {RNNs} for
  acoustic modeling in speech recognition,''
\newblock in {\em ICASSP}, 2017.

\bibitem{zeineldeen:icassp20}
M. Zeineldeen, A. Zeyer, R. Schlüter, and H. Ney,
\newblock ``Layer-normalized {LSTM} for hybrid-{HMM} and end-to-end {ASR},''
\newblock in {\em ICASSP}, 2020.

\bibitem{kanthak:icassp2002}
S. Kanthak and H. Ney,
\newblock ``Context-dependent acoustic modeling using graphemes for large
  vocabulary speech recognition,''
\newblock in {\em IEEE International Conference on Acoustics, Speech, and
  Signal Processing}, Orlando, FL, USA, May 2002.

\bibitem{killer2003grapheme}
M. Killer, S. Stuker, and T. Schultz,
\newblock ``Grapheme based speech recognition,''
\newblock in {\em Eurospeech}, 2003.

\bibitem{sung2009graphemes}
Y.-H. Sung, T. Hughes, F. Beaufays, and B. Strope,
\newblock ``Revisiting graphemes with increasing amounts of data,''
\newblock in {\em ICASSP}. IEEE, 2009.

\bibitem{le2019senones}
D. Le, X. Zhang, W. Zheng, C. F{\"u}gen, G. Zweig, and M.~L. Seltzer,
\newblock ``From senones to chenones: Tied context-dependent graphemes for
  hybrid speech recognition,''
\newblock in {\em ASRU}, 2019.

\bibitem{breiman1984cart}
L. Breiman, J. Friedman, C. Stone, and R. Olshen,
\newblock {\em Classification and regression trees},
\newblock CRC press, 1984.

\bibitem{young1992cart}
S.~J. Young,
\newblock ``The general use of tying in phoneme-based {HMM} speech
  recognisers,''
\newblock in {\em ICASSP}, 1992.

\bibitem{irie19:phoneme}
K. Irie, R. Prabhavalkar, A. Kannan, A. Bruguier, D. Rybach, and P. Nguyen,
\newblock ``On the choice of modeling unit for sequence-to-sequence speech
  recognition,''
\newblock in {\em Interspeech}, 2019.

\bibitem{sak2015fast}
H. Sak, A. Senior, K. Rao, and F. Beaufays,
\newblock ``Fast and accurate recurrent neural network acoustic models for
  speech recognition,''
\newblock in {\em Sixteenth Annual Conference of the International Speech
  Communication Association}, 2015.

\bibitem{miao2016ctc}
Y. Miao, M. Gowayyed, X. Na, T. Ko, F. Metze, and A. Waibel,
\newblock ``An empirical exploration of {CTC} acoustic models,''
\newblock in {\em ICASSP}, 2016.

\bibitem{sainath2018nolex}
T.~N. Sainath, R. Prabhavalkar, S. Kumar, S. Lee, A. Kannan, D. Rybach, V.
  Schogol, P. Nguyen, B. Li, Y. Wu, et~al.,
\newblock ``No need for a lexicon? evaluating the value of the pronunciation
  lexica in end-to-end models,''
\newblock in {\em ICASSP}, 2018.

\bibitem{zhou2018comparison}
S. Zhou, L. Dong, S. Xu, and B. Xu,
\newblock ``A comparison of modeling units in sequence-to-sequence speech
  recognition with the {Transformer} on {Mandarin} {Chinese},''
\newblock in {\em ICONIP}, 2018.

\bibitem{wang2020phonebpe}
W. Wang, Y. Zhou, C. Xiong, and R. Socher,
\newblock ``An investigation of phone-based subword units for end-to-end speech
  recognition,'' Preprint arXiv:2004.04290, 2020.

\bibitem{kubo2020joint}
Y. Kubo and M. Bacchiani,
\newblock ``Joint phoneme-grapheme model for end-to-end speech recognition,''
\newblock in {\em ICASSP}, 2020.

\bibitem{hu2019phoneme}
K. Hu, A. Bruguier, T.~N. Sainath, R. Prabhavalkar, and G. Pundak,
\newblock ``Phoneme-based contextualization for cross-lingual speech
  recognition in end-to-end models,'' Preprint arXiv:1906.09292, 2019.

\bibitem{variani2020hat}
E. Variani, D. Rybach, C. Allauzen, and M. Riley,
\newblock ``Hybrid autoregressive transducer ({HAT}),''
\newblock in {\em ICASSP}, 2020.

\bibitem{watanabe2017ctcatt}
S. Watanabe, T. Hori, S. Kim, J.~R. Hershey, and T. Hayashi,
\newblock ``Hybrid {CTC}/attention architecture for end-to-end speech
  recognition,''
\newblock {\em IEEE Journal of Selected Topics in Signal Processing}, vol. 11,
  no. 8, 2017.

\bibitem{collobert2016wav2letter}
R. Collobert, C. Puhrsch, and G. Synnaeve,
\newblock ``Wav2letter: an end-to-end convnet-based speech recognition
  system,'' Preprint arXiv:1609.03193, 2016.

\bibitem{zhou2021icassp}
W. Zhou, S. Berger, R. Schl\"uter, and H. Ney,
\newblock ``Phoneme based neural transducer for large vocabulary speech
  recognition,''
\newblock in {\em submitted to ICASSP}, 2021.

\bibitem{chorowski2015attention}
J. Chorowski, D. Bahdanau, D. Serdyuk, K. Cho, and Y. Bengio,
\newblock ``Attention-based models for speech recognition,''
\newblock in {\em Advances in neural information processing systems}, 2015.

\bibitem{chan2016las}
W. Chan, N. Jaitly, Q.~V. Le, and O. Vinyals,
\newblock ``Listen, attend and spell: A neural network for large vocabulary
  conversational speech recognition,''
\newblock in {\em ICASSP}, 2016.

\bibitem{hochreiter1997lstm}
S. Hochreiter and J. Schmidhuber,
\newblock ``Long short-term memory,''
\newblock {\em Neural computation}, vol. 9, no. 8, 1997.

\bibitem{zeyer2018:returnn}
A. Zeyer, T. Alkhouli, and H. Ney,
\newblock ``{RETURNN} as a generic flexible neural toolkit with application to
  translation and speech recognition,''
\newblock in {\em ACL}, 2018.

\bibitem{rybach2011:rasr}
D. Rybach, S. Hahn, P. Lehnen, D. Nolden, M. Sundermeyer, Z. T{\"u}ske, S.
  Wiesler, R. Schl{\"u}ter, and H. Ney,
\newblock ``{RASR} - the {RWTH} {Aachen} {University} open source speech
  recognition toolkit,''
\newblock in {\em ASRU}, 2011.

\bibitem{godfrey1992switchboard}
J.~J. Godfrey, E.~C. Holliman, and J. McDaniel,
\newblock ``Switchboard: Telephone speech corpus for research and
  development,''
\newblock in {\em ICASSP}, 1992.

\bibitem{nguyen2019improving}
T. Nguyen, S. Stueker, J. Niehues, and A. Waibel,
\newblock ``Improving sequence-to-sequence speech recognition training with
  on-the-fly data augmentation,'' Preprint arXiv:1910.13296, 2019.

\bibitem{karita2019trafo}
S. Karita, N. Chen, T. Hayashi, T. Hori, H. Inaguma, Z. Jiang, M. Someki, N.
  Soplin, R. Yamamoto, X. Wang, et~al.,
\newblock ``A comparative study on {Transformer} vs {RNN} in speech
  applications,''
\newblock in {\em ASRU}, 2019.

\bibitem{Panayotov2015librispeech}
V. {Panayotov}, G. {Chen}, D. {Povey}, and S. {Khudanpur},
\newblock ``Librispeech: An {ASR} corpus based on public domain audio books,''
\newblock in {\em ICASSP}, 2015.

\end{thebibliography}

\end{document}